\documentclass[fleqn,12pt,twoside]{article}
\usepackage{espcrc1}
\input epsf.sty
\input psfig.sty

\newcommand{\AmS}{{\protect\the\textfont2
  A\kern-.1667em\lower.5ex\hbox{M}\kern-.125emS}}

\title{$P_{t}$ Distributions of Identified Charged Hadrons Measured with
the PHENIX Experiment at RHIC }

\author{J.Velkovska \address[MCSD]{SUNY at Stony Brook} for the PHENIX
        Collaboration %
	\thanks{For the full PHENIX Collaboration author list and acknowledgments see the contribution by W.A. Zajc (K. Adcox {\it et al.}) in this volume.}}
\begin{document}
       

\maketitle

\begin{abstract}
Transverse momentum spectra of identified
$\pi^{+}$, $\pi^{-}$, $K^{+}$, $K^{-}$, p and $\overline{p}$ were
measured by the PHENIX experiment at mid-rapidity in Au-Au collisions
at $\sqrt{s_{NN}}=130$ GeV over a broad momentum range. 
Inverse slope parameters and $\langle p_{t}\rangle $ were
measured in minimum bias events and as a function of the number of
participating nucleons.  The mass and centrality dependence of the
inverse slope parameters and $\langle p_{t}\rangle $ is presented and discussed.

\end{abstract}

\section{Introduction}

With the beginning of RHIC operation in the summer of 2000, a new
regime of heavy-ion collisions at ultra-relativistic energies has
become available. A measurement of the hadron $p_{t}$ distributions gives
important information about the evolution of the system and the
conditions at freeze-out. The particle ratios, obtained from the
transverse momentum spectra, are sensitive to the chemical properties 
of the system and the particle production mechanism \cite{Hiroaki}. 

\section{Set-up and Data Analysis}
The PHENIX detector has various particle identification (PID)
capabilities \cite{Ham}, including  excellent hadron identification
over a broad momentum range. The measurement presented in this paper
was performed using
the Beam-Beam counters (BBC) and the Time-of-flight (TOF) counters as
the timing system, while the 
drift chamber and pad chamber 1 provided  the momentum measurement.
The acceptance covers the pseudo-rapidity region $|\eta|<0.35$ and 
azimuthal angle $\Delta \phi = \pi/4$. Hadron identification was done
in mass$^{2}$ versus momentum space using momentum-dependent PID bands
which take into account the measured TOF and momentum resolution. 
Corrections to the raw spectra were obtained using a single particle
Monte-Carlo simulation. Multiplicity dependent efficiencies were studied by
embedding Monte-Carlo tracks into real events. Track-by-track
efficiency corrections were applied, taking into account the event multiplicity.   

\section{Results and Discussion}

\begin{figure}
\begin{center}
\centerline{\psfig{file=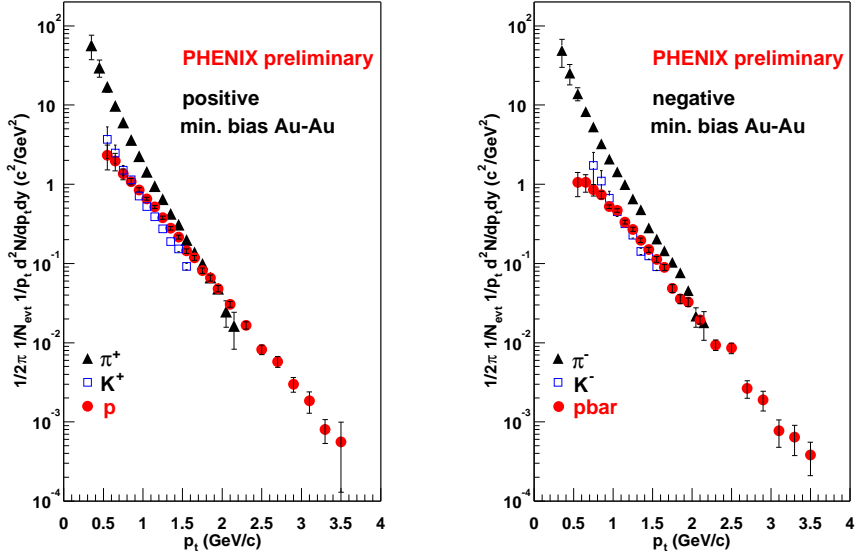,height=8cm,angle=-90}}
\end{center}
\vskip -1.2cm
\caption{Minimum bias transverse momentum distributions for positive
(left) and negative (right) identified hadrons. The error bars include statistical errors and systematic errors in the acceptance and decay corrections. Additional $20 \%$ sytematic errors on the absolute normalization are not included. \label{f:fig1}}
\end{figure}

Figure~\ref{f:fig1} shows the minimum bias $p_{t}$ spectra for
$\pi^{+/-}$, $K^{+/-}$, $p$ and 
$\overline{p}$ . The lower $p_{t}$-cutoffs were imposed due to loss of 
acceptance and tracking efficiency, while the higher  $p_{t}$-cutoff values 
for $\pi^{+/-}$ and $K^{+/-}$ are dictated by the detector capability in $\pi/K$ mass separation.
  
\begin{figure}
\begin{center}
\centerline{\psfig{file=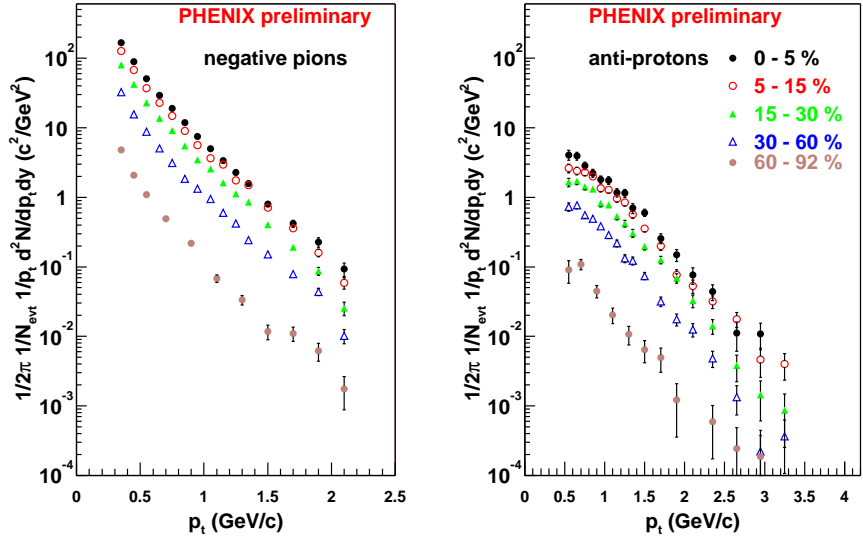,height=8cm,angle=0}}
\end{center}
\vskip -1cm
\caption{Transverse momentum distributions for negative pions (left)
and anti-protons (right) with different centrality selections. The
most central spectra are on the top. The error bars include statistical errors only.\label{f:fig2}}
\end{figure}

The minimum bias data sample was divided into five centrality classes,
using the correlation between the BBC and the Zero-degree
calorimeters \cite{Milov}. The number of
nucleons, participating in the reaction ($N_{part}$),
was obtained using a Glauber model. For each centrality class,
$p_{t}$ distributions were obtained for the six measured particle
species. As an example,  Figure~\ref{f:fig2} shows the
$p_{t}$ distributions for $\pi^{-}$ and $\overline{p}$. The 
centrality selections are indicated in the figure. 

\begin{figure}
\begin{center}
\centerline{\psfig{file=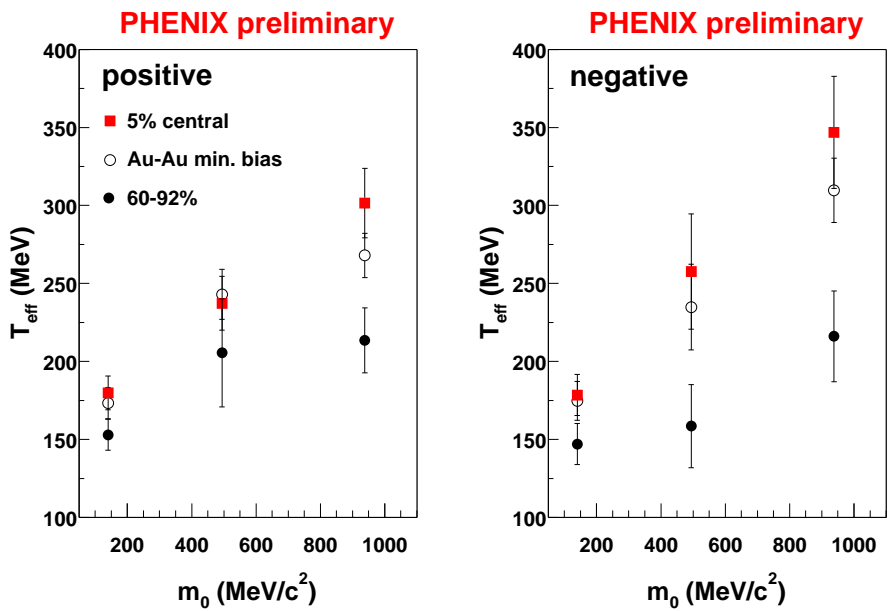,height=8cm,angle=-90}}
\end{center}
\vskip -1.5cm
\caption{Inverse slope parameters as a function of particle mass for
central, minimum bias and peripheral collisions.The error bars include
statistical and systematic errors.\label{f:fig3}}
\end{figure}

The centrality selected transverse mass spectra were fitted with
$a*exp(-m_{t}/T_{eff})$ in the range 0.3 GeV/c $< p_{t} < $ 0.9 GeV/c  for
$\pi^{+/-}$, 0.55 GeV/c $< p_{t} < $ 1.6 GeV/c  for $K^{+}, p $  
$\overline{p}$ and  0.75 GeV/c $< p_{t} < $ 1.6 GeV/c  for $K^{-}$. 
Figure~\ref{f:fig3} shows the inverse slope parameters ($T_{eff}$) as a
function of particle mass for the top 5\% of the total interaction cross-section, minimum bias and
60-92\% central events. A clear mass dependence is seen in the
central and minimum bias events, consistent with radial flow. In 
peripheral collisions, the mass dependence is weaker,
approaching ``$m_{t}$-scaling'' behavior.

\begin{figure}
\begin{center}
\centerline{\psfig{file=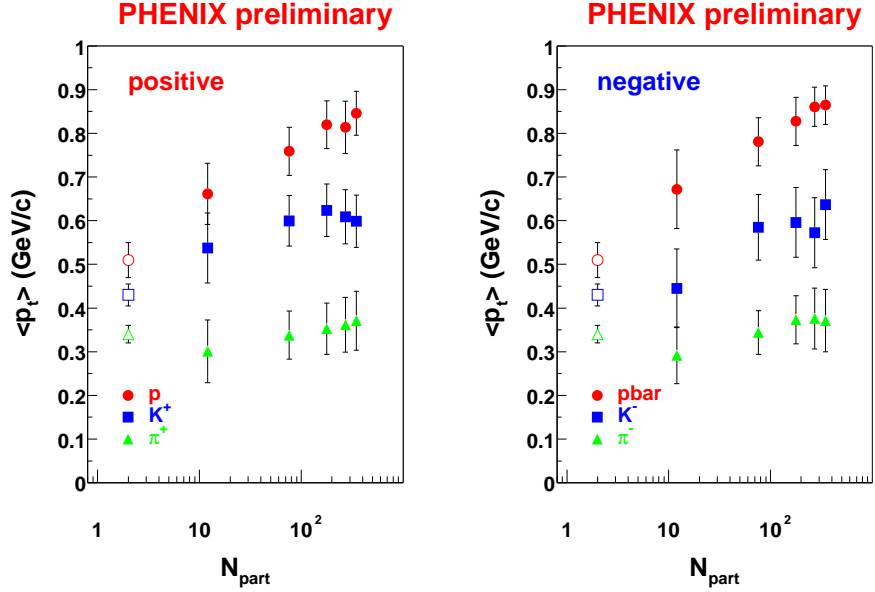,height=8cm,angle=-90}}
\end{center}
\vskip -1cm
\caption{$\langle p_{t}\rangle $ as a function of $N_{part}$ for positive
(left) and negative (right) hadrons. The error bars represent both
statistical and systematic errors. The open points in the figure are
interpolations to $\sqrt{s_{NN}}=130$ GeV from $p\overline{p}$
collisions measured at lower and higher energies \protect\cite{pp}.\label{f:fig4}}
\end{figure}

A complementary measure of the transverse momentum spectra is the mean
$p_{t}$ for each particle species over the whole $p_{t}$ range.
In order to extrapolate the range from 0 to infinity, the $p_{t}$ spectra were
fitted with $a*(p_{0}+p_{t})^{-n}$ for $\pi^{+/-}$ and $a*exp\left(-\sqrt{m^{2}+p_{t}^2}/T\right)$
for $K^{+/-}$, $p$ and $\overline{p}$. The resulting
$\langle p_{t}\rangle $ as a function of $N_{part}$ are plotted in
Figure~\ref{f:fig4}. At RHIC, the pion and kaon $\langle p_{t}\rangle $ is
similar to  that in $p\overline{p}$ collisions, and within errors is independent of
$N_{part}$. The observed $\langle p_{t}\rangle $ for protons
and anti-protons is higher than  in $p\overline{p}$ collisions and seems to increase logarithmically with $N_{part}$.

\section{Conclusion}

The PHENIX measurement of  transverse momentum distributions
of identified charged hadrons at mid-rapidity in Au-Au collisions at
$\sqrt{s_{NN}}=130$ GeV covers a broad momentum range. We have
obtained inverse slope parameters and  $\langle p_{t} \rangle $ for $\pi^{+}$, $\pi^{-}$,
$K^{+}$, $K^{-}$, $p$ and $\overline{p}$ as a function of centrality. 
The inverse slope parameters increase with particle mass, consistent with radial expansion of the system, which increases with centrality. High 
proton and anti-proton inverse slopes persist to high $p_{t}$ leading
to significant baryon yield at
$p_{t}>2$ GeV/c. Baryons may dominate the charged-particle spectrum
beyond this momentum.

\end{document}